# Simultaneous generation of four-wave mixing and stimulated Raman scattering in coupled lithium niobate microdisks mediated by second harmonic generation

Min Wang[1,3], Ni Yao[2], Rongbo Wu[1,3], Zhiwei Fang[1,3,5], Shilong Lv[6], Jianhao Zhang[1,3], Lingling Qiao[1], Jintian Lin[1,†], Wei Fang[2,‡], and Ya Cheng[1,4,7,§]

[1]*State Key Laboratory of High Field Laser Physics, Shanghai Institute of Optics and Fine Mechanics, Chinese Academy of Sciences, Shanghai 201800, China*

[2]*State Key Laboratory of Modern Optical Instrumentation, College of Optical Science and Engineering, Zhejiang University, Hangzhou 310027, China*

[3] *University of Chinese Academy of Sciences, Beijing 100049, China*

[4]*State Key Laboratory of Precision Spectroscopy, East China Normal University, Shanghai 200062, China*

[5]*School of Physical Science and Technology, ShanghaiTech University, Shanghai 200031, China*

[6] *Shanghai Institute of Microsystem and Information Technology, Chinese Academy of Sciences, Shanghai 200050, China*

[7]*Collaborative Innovation Center of Extreme Optics, Shanxi University, Taiyuan, Shanxi 030006, China*

[†]*jintianlin@siom.ac.cn*
[‡]*wfang08@zju.edu.cn*
[§]*ya.cheng@siom.ac.cn*

# Abstract

High quality lithium niobate thin film microresonators provide an ideal platform for on-chip nonlinear optics applications. However, the phase matching condition for efficient parametric process sets a high requirement for fabrication control. Here we demonstrate a photonic molecular structure composed of two strongly coupled lithium niobate microdisks of different diameters fabricated using femtosecond laser micromachining and focused ion beam milling. With a continuous wave excitation, rich nonlinear optical processes including cascaded four-wave mixing and stimulated Raman scattering were observed around the second harmonic wavelength. Our results indicate that the coupled microdisks form a superior system for nonlinear optical process.

The development and commercialization of thin film lithium niobate on insulator (LNOI) has been making a strong impact on photonic integration, which has allowed production of various photonic structures of high optical nonlinearity and electro-optic (EO) tunability in a scalable manner [1,2]. In particular, the LNOI has provided an ideal platform for the fabrication of high quality (high-Q) on-chip lithium niobate (LN) microresonators [3-7] of which the Q factors have approached that of LN microresonators fabricated by mechanical polishing and shaping [8]. Applications of such nonlinear microresonators include classical and quantum light sources [3,5,7,9-15], light modulators [6,16-18], optomechanics [19], and integrated photonic systems [20-22]. Currently, these results are mainly achieved with LN microdisks fabricated using two approaches. The first approach combines femtosecond laser direct writing with focused ion beam (FIB) milling [3,4,23], whilst the second approach utilizes electron-beam lithography (EBL) or photolithography followed by plasma dry etching [5-7]. Both approaches have routinely allowed fabrication of LN microdisk resonators of Q factors on the order of $10^6$. Generally speaking, the first approach, which is established completely based on direct writing processing, offers extreme flexibility for device prototyping. However, the second approach is superior in terms of scalability.

Since the FIB processing can provide a high fabrication resolution down to several nanometers, the first approach mentioned above also facilitates fabrication of delicate structures such as coupled microdisks on LNOI chips. Here, we demonstrated a photonic molecule (PM) structure composed of two LN microdisks with different diameters by precisely controlling the distance between the two microdisks to optimize the coupling strength. Remarkably, we notice that the microdisk PM exhibits unique optical mode features, giving rise to rich nonlinear optical processes including cascaded four-wave mixing (FWM) and stimulated Raman scattering (SRS) around the second harmonic (SH) wavelength. This indicates that PM composed of multiple microdisks of different dimensions offers an effective means for engineering optical mode structures as well as dispersion conditions, both of which are critical for realizing efficient nonlinear optical processes in the LN microresonators.

Figure 1(a) illustrates the three procedures in the fabrication of the LN microdisk PM. In this investigation, we used a commercially available X-cut LN thin film fabricated by ion slicing (NANOLN, Jinan Jingzheng Electronics Co., Ltd) [1]. The LN thin film with a thickness of 700 nm is bonded to a 2 μm thick $SiO_2$ buffer layer coated on a LN wafer. To fabricate the LN microdisk PM, we first patterned the primary structure using femtosecond laser micromachining. Specifically, to minimize the contamination from the debris produced by femtosecond laser ablation, the LN wafer was immersed in water as the water can help prevent the debris from redeposition on the surface of LN disks. The femtosecond laser was focused by an objective lens with a numerical aperture (NA) of 0.8, generating a focal spot of a diameter with ~ 1 μm. The ablation, which began from the top of the LN thin film and ended at a depth of 8 μm, was conducted in a layer by layer manner by setting the layer thickness as 1 μm. The primary structure (i.e., the peanut-shaped structure shown at the left bottom corner of Fig. 1(a)) fabricated by femtosecond laser micromachining was slightly larger than the designed LN microdisk PM, as its size would be further reduced in the FIB milling process. In particular, the two microdisks were connected side by side as the femtosecond laser ablation could not meet the gap fabrication requirement. In the second procedure, a tiny gap between the two microdisks were produced by FIB (Helios NanoLab 600 Dual Beam, FEI, Thermo Fisher Scientific Inc.) milling with a beam current of ~500 pA. The trajectory of FIB milling followed two white circles was to define the boundary of microdisks, as shown in the middle of the bottom row of Fig. 1(a). It is noteworthy that by properly choosing the parameters of the ion beam (e.g., beam current, depth, dwell time, etc.), we can fine tune the distance between the two microdisks to achieve strong coupling between the whispering gallery modes (WGMs) in two microdisks. In the last procedure, the fabricated structure was immersed in a buffered hydrofluoric acid (HF) solution (BUFFER HF IMPROVED, Transene Co., Inc.) to partially remove the $SiO_2$ layer beneath the LN microdisks. By controlling the duration of the wet etching process, two freestanding microdisks supported by the $SiO_2$ pedestals were produced, forming the LN microdisk PM. Figure 1(b) shows the scanning electron microscope (SEM) image of the PM composed of two microdisks with different diameters of 25 μm and 35 μm. The zoom-in SEM image of the edge of the microdisk presented in Fig.1(c) shows the smooth sidewall with a taper angle of approximately 8°34' which is typical in microstructures fabricated with FIB. Figure 1(d) shows typical cross section at the coupling point of microdisk PM cut by FIB after the deposition of a thin layer of platinum (Pt) for protection. And the width of the gap between the two disks is determined to be 126.9 nm.

Figure 2 schematically shows the experimental setup for investigating the nonlinear phenomena in the fabricated LN microdisk PM. Coupling light into and out of the PM was achieved by bringing a 0.9 μm-diameter fiber taper in close proximity to the periphery of the smaller microdisk [24]. The coupling efficiency was adjusted by controlling the relative position between the fiber taper and the microdisk. A swept spectrometer (Model 4650, dBm Optics Inc.) with a built-in narrow-bandwidth continuous-wave tunable diode laser (Model 688-LN, New Focus) was used to characterize the optical mode distribution of the PM. The tunable laser has a linewidth of 10 MHz and a spectral range from 1510 nm to 1620 nm. In order to excite nonlinear processes in the PM, the tunable laser was boosted by an erbium-ytterbium-doped fiber amplifier (EDFA) to serve as the pump source. The optical power and polarization of the pump laser beam were adjusted by a variable optical attenuator (VOA) and an online fiber polarization controller, respectively. A 50: 50 fiber coupler was used to monitor the power of the pump laser. To investigate the nonlinear optical signals generated in the PM, the image of microdisk PM was projected on the entrance slit of a spectrometer (Shamrock SR-303i-B Andor Technology Ltd.) using a 20× objective lens with a numerical aperture of 0.25, and the emission from the outside boundary of the large microdisk was measured by the spectrometer. Two pieces of short-pass dichroic mirrors (DMSP1180, Thorlabs Co.) were used to block the pumping laser during the spectral measurements in the SH wavelength range. To determine the polarization of the nonlinear optical signals, a calibrated wire grid polarizer was inserted between the short-pass filters and the spectrometer. In addition to the spectral analysis, two charge-coupled-device (CCD) cameras were installed above and beside the PM to monitor the nonlinear processes in the PM, respectively, as shown in Fig. 2.

Figure 3(a) shows the transmission spectrum of the X-cut LN microdisk PM near the wavelength of 1560 nm. To obtain the intrinsic Q factor of the PM system, we chose a low power at 200 µW. Due to the coupling between the two microdisks, the WGM mode structure in the PM appears more complex than that typically observed in single LN microdisk resonators [3-6,9,12]. Four modes were chosen for characterizing the Q factors, as indicated by the red, green, blue and purple lines in Fig. 3. Lorentz fitting curves of these troughs indicate that the resonant modes at 1559.11 nm, 1559.36 nm, 1560. 074 nm, and 1560.515 nm have Q factors of $2.63 \times 10^5$, $1.79 \times 10^5$, $6.37 \times 10^5$, and $8.13 \times 10^5$, respectively.

When the pump laser power increased, strong nonlinear optical phenomena were observed at certain excitation wavelengths. The interference patterns of the strong SH waves are clearly visible along the peripheries of both the microdisks, as shown in the Fig. 4(a). This confirms that under such a condition, the SH wave were in resonance with coupled modes in the PM. The spectra near the SH wavelength collected from the large microdisk at various pump powers are shown in Fig. 4(b). When the pump power was set at 10.4 mW, only the SH peak at 782.73 nm was detected, and polarization measurement confirmed that it was quasi-transverse electric (TE) mode. Surprisingly, when the laser pump power was increased, new peaks started to appear in the vicinity of SH peak as evidenced in Fig. 4(b). When the pump power was 14 mW, two equally spaced (4.6 nm) FWM peaks emerged symmetrically around the SH mode, with wavelengths at 778.09 nm (F1) and 787.37 nm (F2), respectively. As revealed by Fig.

4(b) and (c) that with increase of the pump power, a cascaded FWM process occurred leading to generation of the waves peaked at 773.44 nm (F3) and 792.01 nm (F4). It is noteworthy that the mode spacing of 4.6 nm corresponds to twice of the free spectral range (FSR) of fundamental WGMs in the large microdisk. The FSR of 2.3 nm was calculated based on finite-difference time-domain method. As indicated by the red arrow in the inset of Fig. 4(c), there is a weak signal wave peaked 2.3 nm away from the central wavelength of SH wave which is merely visible from the broadened spectrum of the SH wave. The broadening of the SH spectrum is a result of the mode coupling between the two microdisks, which leads to the transfer of the FWM signal wave from the large disk to the smaller one. In such a case, the loss at the FWM signal wavelength will increase, leading to the less efficient parametric amplification of the signal as well as the idler waves for these WGMs. For the WGMs spectrally separated further away from the SH wave, they can preserve a high Q factor to ensure the efficient cascaded FWM processes with the mode spacing of 4.6 nm. The summed power of the FWM signals (i.e., F1 and F2) as a function of the pump laser power was measured and plotted in Fig. 4(d). The threshold power of FWM in the LN microdisk PM was determined to be 11.07 mW, above which the power of FWM signals increases linearly with the fourth power of the pump laser power at 1565.46 nm wavelength. From the slope of the fitting line in Fig. 4(d), a normalized conversion efficiency of $1.49 \times 10^{-8}$ $mW^{-3}$ can be determined. In other words, we can obtain the FWM signals with a total power of 2.38 µW at a pump power of 20 mW. For comparison, the highest conversion efficiency of the SH generation was experimentally determined to be $1.106 \times 10^{-3}$ $mW^{-1}$, which means a SH signal with a power of 442.4 µW was generated at the pump laser power of 20 mW. Meanwhile, at the wavelengths farther away from the SH wavelength, cascaded Raman scattering was observed. The vibrational modes involved in the Raman scattering were 578 $cm^{-1}$, 581 $cm^{-1}$ and 252 $cm^{-1}$, as indicated by the Raman peaks R1, R2 and R3 in the spectrum of Fig. 4(c), respectively. The high Stokes wave intensities (comparing to the pump intensity at SH wavelength) indicate the occurrence of stimulated Raman scattering [25,26].

Since all the nonlinear spectral measurements mentioned above are carried out by collecting the signals from the large disk whilst the pump laser is coupled into the PM from the small disk, it is unambiguous that the two disks are efficiently coupled to each other. The top view image in Fig. 4(a) also provides a solid proof that mode coupling between the two microdisks has been achieved in the microdisk PM around the SH wavelength. In addition, based on the fact that the SH wave at 782.73 nm wavelength is polarized perpendicularly to the pump wave at 1565.46 nm wavelength, it is expected that the high conversion efficiency in SH generation should be a result of the so-called cyclic phase matching (CPM)[9,27], which allows for periodic transient phase matching in the X-cut microdisks as far as the constant effective refractive index of the quasi-transverse magnetic (TM) pump laser can intersect the oscillating effective refractive index of the quasi-TESH wave. In this manner, the CPM provides more flexibility in achieving the phase matched nonlinear optical conversion than the perfect phase matching mechanism [9,10].

One striking observation in our experiment is that the cascaded FWM and SRS are both realized around the SH but not the fundamental wavelength even though the fundamental wave is much stronger. We attribute this to the several facts as follows. First, the short wavelength of the SH wave leads to the high mode density in the microdisks. In addition, when the modes in the microdisks are coupled to each other, mode broadening and splitting can occur. The combination of the high mode density and the mode splitting near ~780 nm wavelength makes it easy to realize the multiple resonance condition at the wavelengths that also satisfies the energy conservation relation $2\omega_{SH}=\omega_{F1}+\omega_{F2}$. Here $\omega_{SH}$, $\omega_{F1}$, and $\omega_{F2}$ are frequencies of SH, F1, and F2, respectively. Second, we notice that the measured SH wave possesses a quasi-TE mode which is of an oscillating refractive index along the periphery of the large disk. Meanwhile, the FWM signals F1 and F2 both possess quasi-TM modes for which the refractive indices are constant in the microdisk. Under this condition, the CPM can be achieved again. When the multiple resonance and phase matching conditions are both fulfilled, it is straightforward to boost the efficiencies of the nonlinear processes. Since the phase matching condition is automatically ensured for the SRS in the WGM microresonators, it is less critical to generate the Raman lasing in the microresonators as has been reported by many research groups before [7,25,26]. Thus, all the Raman lasing signals are generated by the assistance of the cavity enhanced field of the SH wave in the coupled modes belonging to the microdisk PM system.

To conclude, we have fabricated an LN microdisk PM structure composed of two microdisks with the different diameters. We show that various types of nonlinear processes can be efficiently excited in the PM at once including the SH generation, and SH mediated FWM and SRS under the continuous wave excitation. The nonlinear optical platform provided by the LN microdisk PM can be of great use for generating quantum and classical light sources.

This work is supported by the National Basic Research Program of China (Grant No. 2014CB921303), National Natural Science Foundation of China (Grant Nos. 11734009, 61590934, 61635009, 61327902, 61505231, 11604351, 11674340, 61575211, 61675220), the Strategic Priority Research Program of Chinese Academy of Sciences (Grant No. XDB16000000), Key Research Program of Frontier Sciences, Chinese Academy of Sciences (Grant No. QYZDJ-SSW-SLH010), the Project of Shanghai Committee of Science and Technology (Grant 17JC1400400), Shanghai Rising-Star Program (Grant No. 17QA1404600), and the Fundamental Research Funds for the Central Universities.

M. W. and N. Y. contributed equally to this work.

**Captions of figures:**

Fig. 1 (Color online) (a) Fabrication procedures of a photonic molecule composed of two LN microdisks: water-assisted femtosecond laser ablation, focused ion beam (FIB) milling and hydrofluoric acid wet etching. The black arrow shows the optic axis of the LN thin film. (b) The SEM image of the LN microdisk PM. (c) The zoom-in SEM image of the edge of the large microdisk, indicating a freestanding disk with a smooth sidewall. (d) The SEM image of the cross section at the coupling point of the PM cut by FIB with a gap size as small as 126.9 nm.

Fig. 2 (Color online) Experimental setup for investigating the nonlinear processes in LN microdisk PM.

Fig. 3 (Color online) (a) A typical mode distribution spectrum of the PM. The Lorentz fitting curves of the peaks colored in red, green, blue and purple are shown in (b)-(e), indicating a Q factor of $2.63 \times 10^5$, $1.79 \times 10^5$, $6.37 \times 10^5$, and $8.13 \times 10^5$, respectively.

Fig. 4 (Color online) (a) Optical micrograph of the microdisk PM when generating the SH wave. (b) Nonlinear spectra generated near the SH wavelength at the pump powers of 10.4 mW, 14 mW, 17.1 mW and 21.9 mW. (c) Spectrum at the pump power of 23.2 mW, left inset: detailed FWM peaks, right inset: spectrum near the pump wavelength, showing that there is no visible nonlinear peaks in vicinity. (d) Power of the FWM as a function of the forth power of pump power. .

Fig. 1

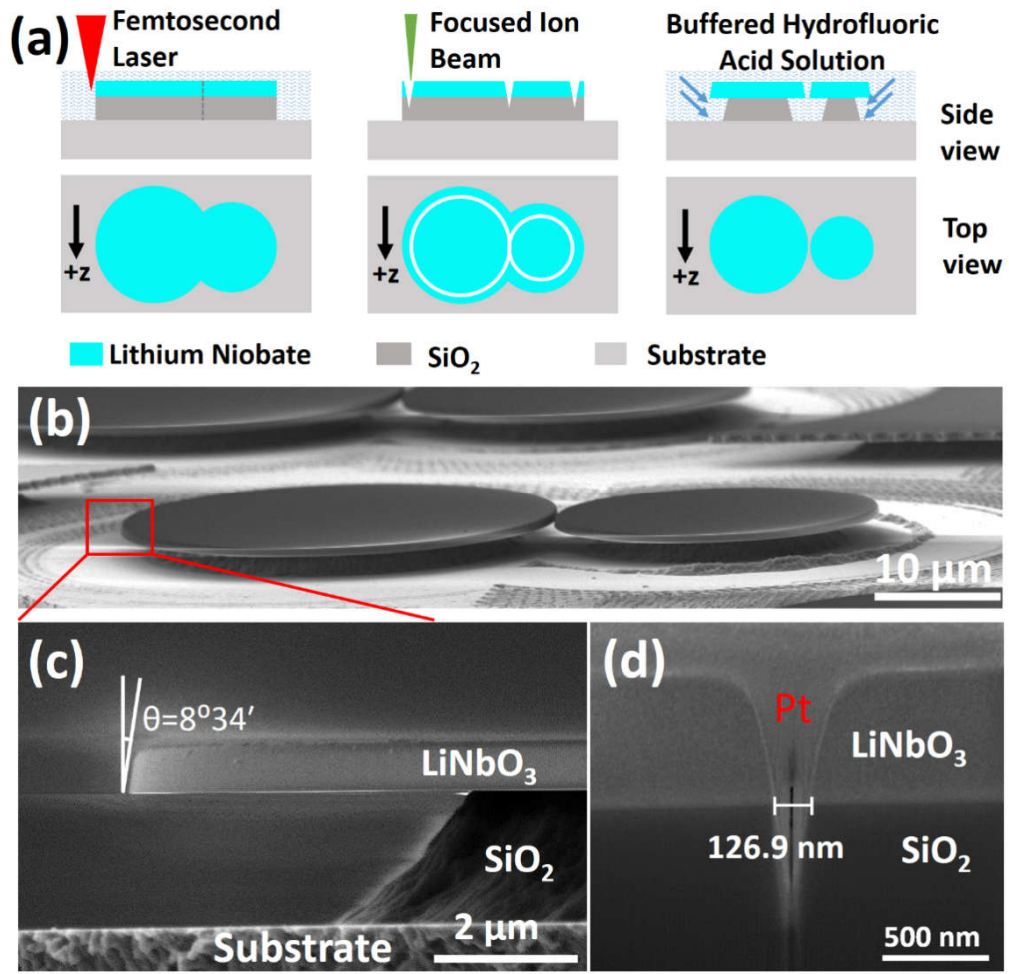

Fig. 2

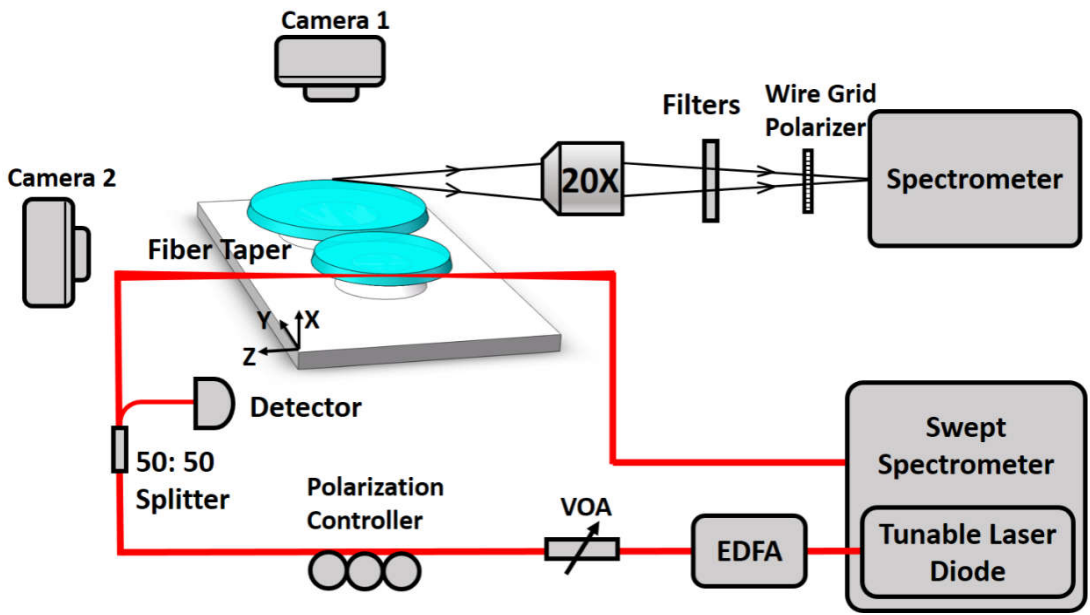

Fig. 3

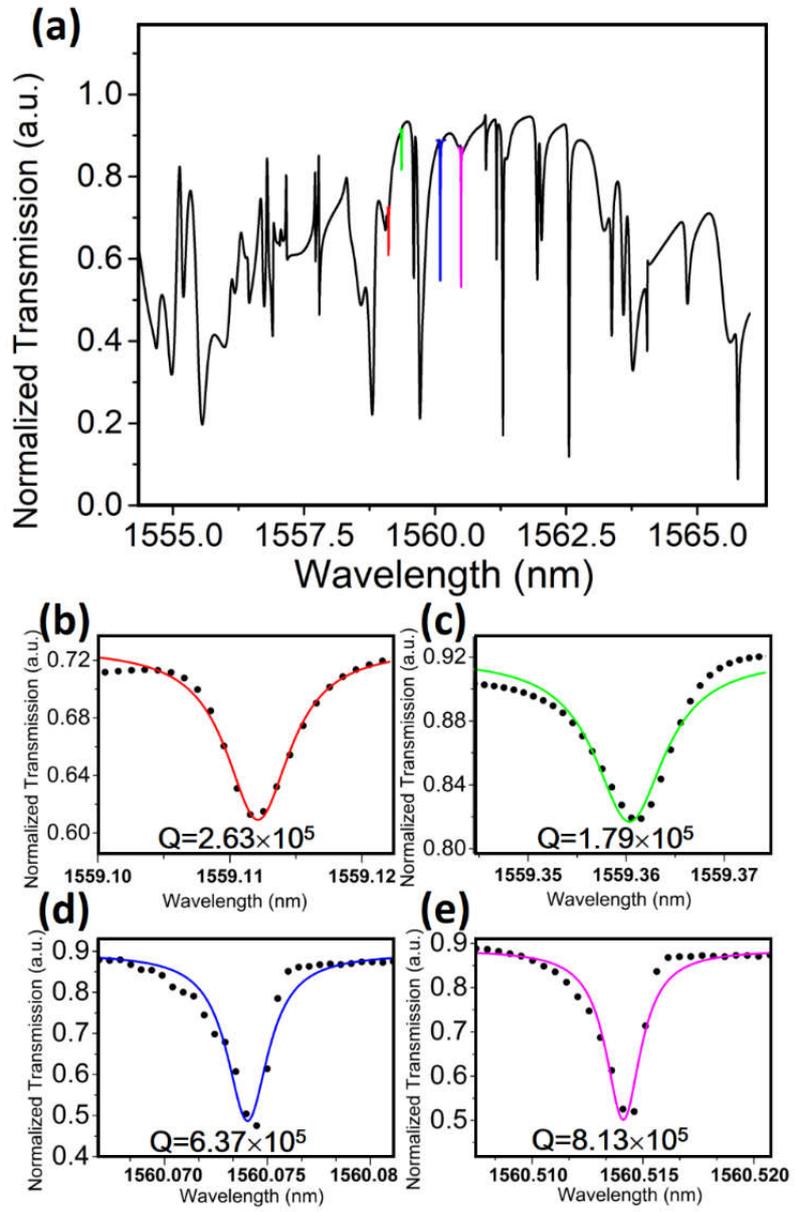

Fig. 4

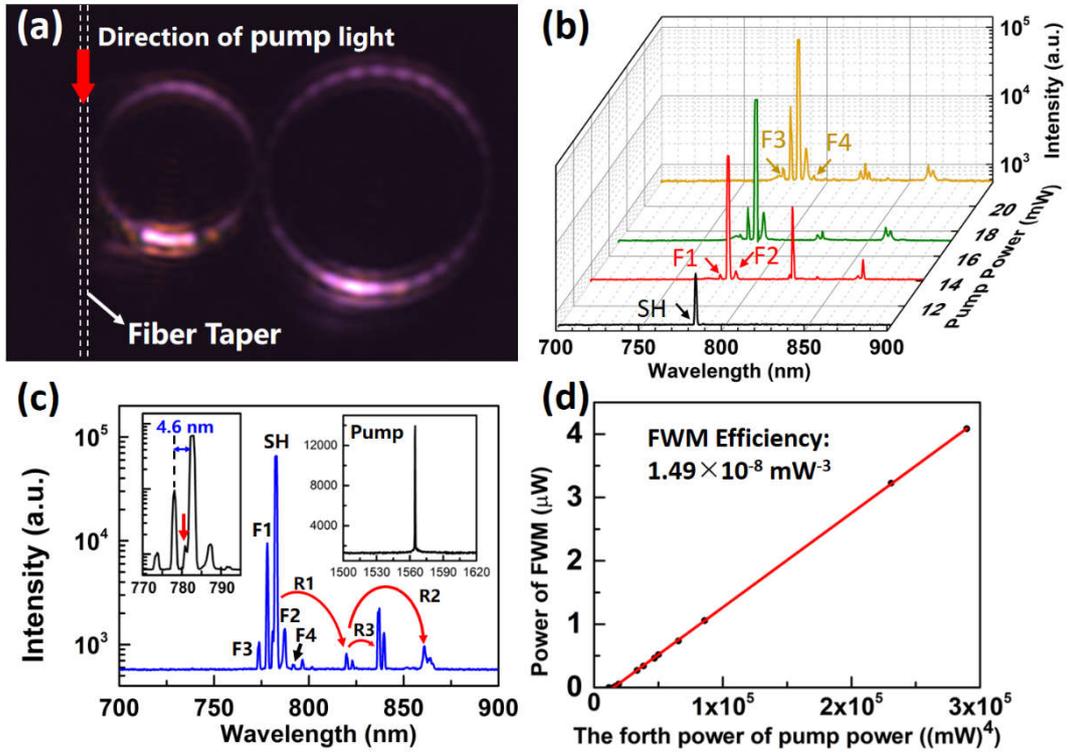